\documentclass{article}
\usepackage{amsmath}
\usepackage{cite}
\usepackage{graphicx}
\usepackage{dcolumn}

\begin{document}

\date{}
\title{On the eigenvalues of the harmonic oscillator with a Gaussian perturbation}
\author{Paolo Amore\thanks{%
e--mail: paolo@ucol.mx} \\
Facultad de Ciencias, CUICBAS, Universidad de Colima,\\
Bernal D\'{\i}az del Castillo 340, Colima, Colima,Mexico \\
Francisco M. Fern\'andez\thanks{%
e--mail: fernande@quimica.unlp.edu.ar} \\
INIFTA, Divisi\'{o}n Qu\'{\i}mica Te\'{o}rica,\\
Blvd. 113 y 64 (S/N), Sucursal 4, Casilla de Correo 16,\\
1900 La Plata, Argentina \\
Javier Garcia \\
Instituto de F\'{i}sica La Plata, CONICET\\
and Departamento de F\'{i}sica, UNLP,\\
C.C. 67, 1900 La Plata, Argentina}
\maketitle

\begin{abstract}
We test the analytical expressions for the first two eigenvalues
of the harmonic oscillator with a Gaussian perturbation proposed
recently. Our numerical eigenvalues show that those expressions
are valid in an interval of the coupling parameter that is greater
than the one estimated by the authors. We also calculate critical
values of the coupling parameter and several exceptional points in
the complex plane.
\end{abstract}

\section{Introduction}

\label{sec:intro}

Fassari et al\cite{FNR20} studied a harmonic oscillator with a Gaussian
perturbation. By means of the Fredholm determinant they derived analytical
expressions for the first two eigenvalues. These expressions resemble those
commonly obtained by perturbation theory as they are polynomials of second
degree in the coupling constant. They showed a figure for these approximate
eigenvalues as functions of the coupling constant but did not compare their
results with any independent calculation. The purpose of this paper is to
test the accuracy of those results and derive some relevant mathematical
properties of the eigenvalues of the model.

In section~\ref{sec:analytical} we introduce the model and show some
analytical results including the polynomial expressions of Fassari et al\cite
{FNR20}. In section~\ref{sec:num_app} we briefly discuss two methods for the
calculation of the eigenvalues and related quantities. In section~\ref
{sec:critical} we calculate some critical values of the coupling constant.
In section~\ref{sec:EPS} we obtain the exceptional points of the eigenvalues
of the model. Finally, in section~\ref{sec:conclusions} we summarize the
main results and draw conclusions.

\section{Analytical expressions}

\label{sec:analytical}

The Hamiltonian operator in the coordinate representation is
\begin{equation}
H=H_{0}+\lambda V,\;H_{0}=-\frac{1}{2}\frac{d^{2}}{dx^{2}}+\frac{1}{2}%
x^{2},\;V(x)=-e^{-x^{2}}.  \label{eq:H}
\end{equation}
Fassari et al\cite{FNR20} considered the cases $\lambda >0$ and $\lambda <0$
separately but it is convenient to treat them in a unified way. Since the
potential is parity invariant the solutions $\psi (x)$ to the
Schr\"{o}dinger equation $H\psi =E\psi $ are either even ($\psi (-x)=\psi (x)
$) or odd ($\psi (-x)=-\psi (x)$) so that we can treat the two symmetry
species separately. The eigenvalues $E(\lambda )$ of $H$ decrease with $%
\lambda $ as predicted by the Hellmann-Feynman theorem\cite{G32,F39}
\begin{equation}
\frac{dE}{d\lambda }=-\left\langle e^{-x^{2}}\right\rangle .  \label{eq:HFT}
\end{equation}

By means of a somewhat laborious method Fassari et al\cite{FNR20} obtained
\begin{eqnarray}
E_{0}^{PT} &=&\frac{1}{2}-\frac{\lambda }{\sqrt{2}}-\frac{\ln {\left( 8-4%
\sqrt{3}\right) }}{2}\lambda ^{2},  \nonumber \\
E_{1}^{PT} &=&\frac{3}{2}-\frac{\sqrt{2}}{4}\lambda -\frac{2\sqrt{3}-3\left[
1-\ln {\left( 8-4\sqrt{3}\right) }\right] }{24}\lambda ^{2},
\label{eq:E0,E1_PT}
\end{eqnarray}
for the two lowest eigenvalues. These expressions show the first two terms
of the perturbation expansions\cite{F01}
\begin{equation}
E_{n}=\sum_{j=0}^{\infty }E_{n,j}\lambda ^{j},  \label{eq:E_n_PT}
\end{equation}
for $n=0$ and $n=1$ (even and odd solutions, respectively). According to the
HFT (\ref{eq:HFT}) $E_{n,1}<0$, $n=0,1,\ldots $ and the standard equations
of perturbation theory\cite{F01} show that $E_{n,2}<0$ for $n=0$ and $n=1$
in agreement with equation (\ref{eq:E0,E1_PT}). The perturbation series (\ref
{eq:E_n_PT}) in general and the analytical expressions (\ref{eq:E0,E1_PT})
in particular are valid for all values of $\lambda $ within its radius of
convergence. This point is discussed below in section~\ref{sec:EPS}.

\section{Numerical approaches}

\label{sec:num_app}

The calculation of perturbation terms of greater order can be rather
cumbersome. For this reason we resort to two numerical methods.

The first one is the well-known Rayleigh-Ritz (RR) variational method that
is based on matrix representations of the Hamiltonian operator in a suitable
orthonormal basis $\left\{ \left| i\right\rangle ,\;i=0,1,\ldots \right\} $%
\cite{M33,F22}. This approach is useful if we can obtain the matrix elements
$H_{i,j}=\left\langle i\right| H\left| j\right\rangle $ in order to build
matrix representations of the form $\mathbf{H}_{D}=\left( H_{i,j}\right)
_{i,j=0}^{D-1}$. The approximate eigenvalues are given by the roots of the
secular determinant $\left| \mathbf{H}_{D}-E\mathbf{I}_{D}\right| =0$, where
$\mathbf{I}_{D}$ is the $D\times D$ identity matrix. The advantage of this
approach is that the approximate eigenvalues $E_{n}^{[D]}(\lambda )$
converge towards the exact ones $E_{n}(\lambda )$ from above: $%
E_{n}^{[D]}(\lambda )\geq E_{n}^{[D+1]}(\lambda )\geq E_{n}(\lambda )$\cite
{M33,F22}. In the present case we may resort to the well-known basis set of
eigenfunctions of $H_{0}$ and, as argued above, apply the method to even and
odd states separately.

The second approach is the Riccati-Pad\'{e} method (RPM)\cite{FMT89a} that
has proved to be quite accurate in the past\cite{FG17,FG18} and that is easy
to apply to this kind of problems. The starting point of this approach is
the logarithmic derivative of the wavefunction $f(x)=s/x-\psi ^{\prime
}(x)/\psi (x)$, where the prime stands for the derivative with respect to $x$
and $s=0$ for even states and $s=1$ for odd ones. The logarithmic derivative
can be expanded in a Taylor series about the origin $f(x)=f_{0}x+f_{1}x^{3}+%
\ldots +f_{j}x^{2j+1}+\ldots $ and the approximate eigenvalues are roots of
the Hankel determinants $H_{D}^{d}(E,\lambda )=\left| f_{i+j+d-1}\right|
_{i,j=1}^{D}$, where $D=2,3,\ldots $ is the dimension of the determinant and
$d$ is related to the Pad\'{e} approximant used in its construction\cite
{FMT89a,FG17,FG18}.

With both methods we obtain the approximate eigenvalues from the roots of
polynomial functions $F_{D}(E,\lambda )$ of $E$ and $\lambda $. It is clear
that we can have either $E(\lambda )$ or $\lambda (E)$.

\begin{figure}[tbp]
\begin{center}
\includegraphics[width=9cm]{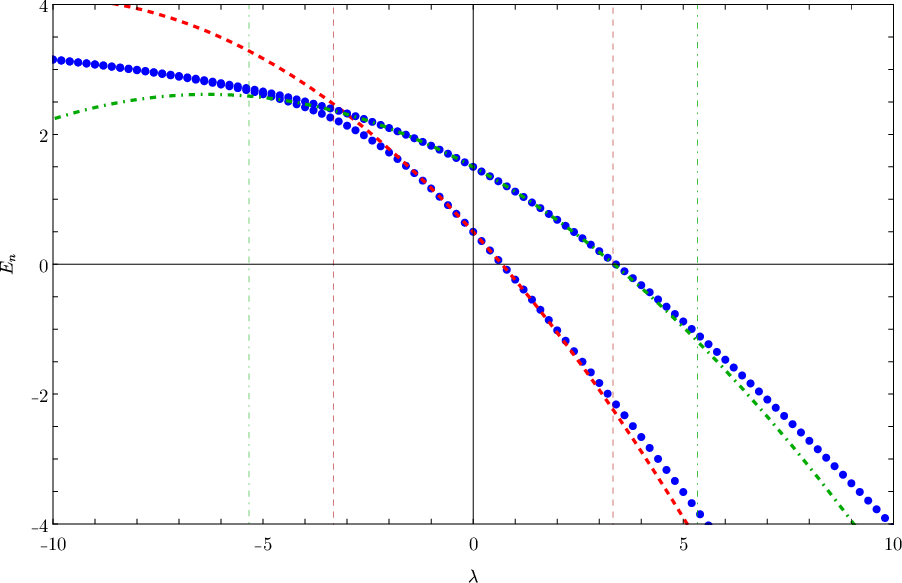}
\end{center}
\caption{Eigenvalues $E_0$ and $E_1$ calculated by means of equations (\ref
{eq:E0,E1_PT}) (red dash and green dash-dot lines) and RR variational method
(blue dots)}
\label{Fig:En}
\end{figure}

Figure~\ref{Fig:En} shows the analytical expressions of Fassari et al\cite
{FNR20} and present numerical results in the interval $-10<\lambda <10$. We
appreciate that $E_{0}^{PT}$ and $E_{1}^{PT}$ are reasonably accurate in the
intervals $-3<\lambda <3$ and $-5<\lambda <5$, respectively. A rigorous
discussion of this point will be given in section~\ref{sec:EPS}. Note that $%
E_{0}(\lambda )$ and $E_{1}(\lambda )$ approach each other as $\lambda $
decreases. The reason is that the potential is a double well when $\lambda
<0 $ and that the barrier between the wells increases as $\lambda $
decreases. As the barrier increases pairs of eigenvalues become
quasi-degenerate as illustrated in figure~\ref{Fig:En} for the two lowest
ones.

\section{Critical values of $\lambda $}

\label{sec:critical}

For $\lambda =0$ all the eigenvalues are positive. Since they decrease with $%
\lambda $ there may be a value $\lambda _{n}^{c}>0$ such that $E\left(
\lambda _{n}^{c}\right) =0$ for a given value of $n$. The methods introduced
above are suitable for this calculation because it is only necessary to
obtain the roots of $F_{D}(0,\lambda )$ for increasing values of $D$.

In the case of the RR we obtain $\lambda _{n}^{c}(D)$ such that $%
E_{n}^{[D]}\left( \lambda _{n}^{c}(D)\right) =0$. Since $E_{n}^{[D]}\left(
\lambda _{n}^{c}(D)\right) \geq E_{n}^{[D+1]}\left( \lambda
_{n}^{c}(D)\right) $ we conclude that $\lambda _{n}^{c}(D+1)\leq \lambda
_{n}^{c}(D)$ (the RR results already exhibit this behaviour). A similar
calculation can be carried out with the RPM but this approach does not
produce bounds.

From equation (\ref{eq:E0,E1_PT}) we obtain $\lambda _{0}^{c}(PT)=$ $0.684$
and $\lambda _{1}^{c}(PT)=3.35$ that agree quite well with the numerical
results $\lambda _{0}^{c}=$ $0.6863528514$ and $\lambda _{1}^{c}=3.393856454$%
. This fact is not surprising because these values of $\lambda _{n}^{c}$ lie
within the range of validity of the analytical polynomial expressions as
shown in figure~\ref{Fig:En}. In section~\ref{sec:EPS} we provide a more
rigorous analysis.

The RPM is suitable for the calculation of extremely accurate results; for
example, in this case we obtained
\begin{eqnarray}
\lambda _{0}^{c} &=&0.686352851432136232145426692879870945398714397832
\nonumber \\
&&4894550439671729746007959840101881351645364483292962  \nonumber \\
\lambda _{1}^{c} &=&3.3938564542892053249137531899771358228367841679214
\nonumber \\
&&716395950274391736560573854993368554164090084557542
\label{eq:lambda_n^c_RPM}
\end{eqnarray}
with $D\leq 380$.

\section{Exceptional points}

\label{sec:EPS}

The radius of convergence of the perturbation series (\ref{eq:E_n_PT}) is
determined by the singularity of $E(\lambda )$ closest to the origin of the
complex $\lambda $-plane. We can estimate branch points of order two in more
than one way. A simple one is based on the discriminant $p_{N}(\lambda
)=Disc_{E}\left( F_{D}(E,\lambda )\right) $, where $p_{N}(\lambda )$ is a
polynomial of degree $N$\cite{AF21} (and references therein). This kind of
singularities is given by the roots of $d\lambda /dE=0$. Since $%
dF_{D}/dE=\partial F_{D}/\partial E+\left( \partial F_{D}/\partial \lambda
\right) \left( d\lambda /dE\right) =0$ then we can obtain the singular
points by solving the set of equations $\left[ F_{D}(E,\lambda )=0,\partial
F_{D}(E,\lambda )/\partial E=0\right] $ by means of a root finder like the
Newton-Raphson method. In the present paper we use the singular points
obtained from the discriminant of the secular determinant provided by the RR
method\cite{AF21} as starting points for the application of the
Newton-Rapson method to the set of equations derived from Hankel
determinants. The singularities of $E(\lambda )$ are commonly called
exceptional points (EPS) (see, for example\cite{AF21} and the references
therein). If $\lambda _{n}^{EP}$ is the exceptional point of $E_{n}(\lambda
) $ closest to the origin of the complex $\lambda $-plane, then the
perturbation series (\ref{eq:E_n_PT}) converges for $|\lambda |<\left|
\lambda _{n}^{EP}\right| $.

\begin{figure}[tbp]
\begin{center}
\includegraphics[width=9cm]{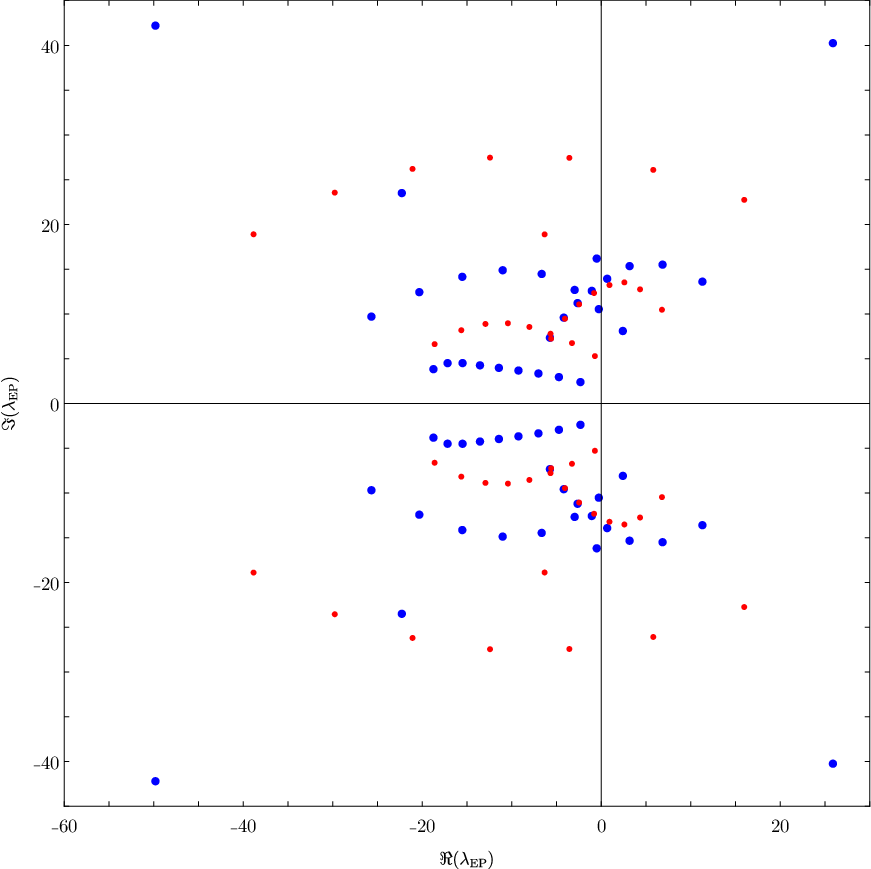}
\end{center}
\caption{Exceptional points for the even (larger blue dots) and odd (smaller
red dots) solutions}
\label{Fig:EPS}
\end{figure}

Figure~\ref{Fig:EPS} shows some of the EPS calculated by means of the RR
variational method and the discriminant as discussed in an earlier paper\cite
{AF21}. By means of the RPM with $D\leq 100$ we obtained the following
highly accurate results
\begin{eqnarray}
\lambda _{0}^{EP} &=&-2.3226516328467993361294340651810891666  \nonumber \\
&&\pm 2.38626692172532053097268204019404075331i  \nonumber \\
\lambda _{1}^{EP} &=&-0.70816242676853914857694137530919271348  \nonumber \\
&&\pm 5.287743791236289699512306550846226281i  \label{eq:lambda_1,2_EP}
\end{eqnarray}
for the two lowest eigenvalues. In principle, such accuracy is unnecessary
for most applications; however, this kind of highly accurate results may be
of utility for testing the convergence of new approaches. The radius of
convergence for the ground state $\left| \lambda _{0}^{EP}\right|
=3.330012076$ and first excited state $\left| \lambda _{1}^{EP}\right|
=5.334953460$ clearly show that the perturbation series (\ref{eq:E_n_PT})
are valid for greater intervals of $\lambda $ than those estimated by
Fassari et al\cite{FNR20}. It is clear that present values of the radii of
convergence of the perturbation series provide a rigorous explanation of the
results in figure~\ref{Fig:En} that were discussed in section~\ref
{sec:num_app} . Note that figure~\ref{Fig:En} shows four vertical lines
located at $\pm \left| \lambda _{0}^{EP}\right| $ and $\pm \left| \lambda
_{1}^{EP}\right| $ to facilitate the analysis.

\section{Conclusions}

\label{sec:conclusions}

We have argued that the analytical expressions derived by Fassari et al\cite
{FNR20} provide the two lowest corrections of perturbation theory for the
two lowest eigenvalues of the harmonic oscillator with a Gaussian
perturbation. By means of two independent numerical methods we calculated
some critical values of the coupling parameter and also the exceptional
points of the eigenvalues of the model. The latter quantities enabled us to
determine that the range of validity of the power series (\ref{eq:E0,E1_PT})
is greater that the one estimated by the authors.

\end{document}